\documentclass[12pt]{article}
\usepackage{amsmath}
\usepackage{graphicx,psfrag,epsf}
\usepackage{enumerate}
\usepackage{natbib}
\usepackage{algorithm}
\usepackage{algorithmic}
\usepackage{epsfig}
\usepackage{subfigure}
\usepackage{amssymb}
\usepackage{setspace}

\usepackage{url} 

\newcommand{\blind}{1}

\begin{document}

\def\spacingset#1{\renewcommand{\baselinestretch}%
{#1}\small\normalsize} \spacingset{1}

\doublespacing


\if1\blind
{
  \title{Bayesian analysis of three parameter singular Marshall-Olkin bivariate Pareto distribution}
  \author{Biplab Paul, Arabin Kumar Dey and Debasis Kundu \\
    Department of Mathematics, IIT Guwahati\\ Department of Mathematics, IIT Guwahati\\   
    Department of Mathematics and Statistics, IIT Kanpur} 

\maketitle

} \fi

\if0\blind
{
  \bigskip
  \bigskip
  \bigskip
  \begin{center}
    {\LARGE\bf Bayesian analysis of three parameter singular Marshall-Olkin bivariate Pareto distribution}
\end{center}
  \medskip
} \fi

\bigskip
\begin{abstract}

 This paper provides bayesian analysis of singular Marshall-Olkin bivariate Pareto distribution.  We consider three parameter singular Marshall-Olkin bivariate Pareto distribution.  We consider two types of prior - reference prior and gamma prior.  Bayes estimate of the parameters are calculated based on slice cum gibbs sampler and Lindley approximation.  Credible interval is also provided for all methods and all prior distributions.  A data analysis is kept for illustrative purpose.   

\end{abstract}

\noindent%
\vfill

\newpage
\spacingset{1.45} 
\section{Introduction}
\label{intro}
 
 Bivariate Pareto distribution (BVPA) is used in modelling data related to climate, network-security etc.  A variety of bivariate (multivariate) extensions of the Pareto distribution also have been considered in the literature. These include the distributions of \cite{SankaranKundu:2014}, \cite{Yeh:2000}, \cite{Yeh:2004}, \cite{AsimitFurmanVernic:2010}.

  In this paper we consider a special type of bivariate Pareto distribution, namely Marshall-Olkin bivariate Pareto (MOBVPA) whose marginals are type-II univariate Pareto distributions.  We use the notation MOBVPA for singular version of this bivariate Pareto.  Finding efficient estimation technique to estimate the parameters of BVPA was a major challenge for last few decades.  The problem is attempted by some authors in frequentist set up through EM algorithm [\cite{AsimitFurmanVernic:2016}, \cite{DeyPaul:2017}].  There is no work in bayesian set up for  singular Marshall-Olkin bivariate Pareto distribution.  In this paper we restrict ourselves only up to three parameter MOBVPA.  

 The bayes estimator can not be obtained in closed form.  Therefore we propose to use two methods. (1) Lindley approximation [\cite{Lindley:1980}] (2) Slice cum Gibbs Sampler [\cite{Neal:2003}, \cite{CasellaGeorge:1992}].  However we can use other Monte Carlo methods for the same.  In this paper we made slight modification in calculation of the Lindley approximation.  We use EM algorithms instead of MLE.  We also calculate credible intervals for the parameters.  Bayes estimators exist even when MLEs do not exist. Also Bayesian estimators may work reasonably well with suitable choice of prior even when MLE's performance is extremely poor.  Therefore working in bayesian set up with such a complicated distribution has its own advantages.  In this paper both informative prior like gamma prior and non-informative prior like reference prior is used.   

 Rest of the paper is organized as follows.  In section 2, we show the bayesian analysis of singular Marshall-Olkin bivariate Pareto distribution.  Numerical results are discussed in section 3. In section 4, A data analysis is shown for illustrative purpose.  We conclude the paper in section 5.        

\section{Bayesian Analysis of singular Marshall-Olkin bivariate Pareto distribution}

 A random variable X is said to have Pareto of second kind, i.e.\\ $X \sim Pa(II)(\mu, \sigma, \alpha)$ if it has the survival function  
$$ \bar{F}_{X}(x ; \mu, \sigma, \alpha) = P(X > x) = (1 + \frac{x - \mu}{\sigma})^{-\alpha} $$
and the probability density function (pdf) $$ f(x ; \mu, \sigma, \alpha) = \frac{\alpha}{\sigma}(1 + \frac{x - \mu}{\sigma})^{-\alpha - 1} $$
with $x > \mu \in \mathcal{R}$, $\sigma > 0$ and $\alpha > 0$.

 Let $U_0, U_1$ and $U_2$ are mutually independent random variable where $U_0 \sim PA(II)(0, 1, \alpha_0)$, $U_1 \sim PA(II)(\mu_1, \sigma_1,\alpha_1)$ and  $U_2 \sim PA(II)(\mu_2, \sigma_2, \alpha_2)$. We define $X_1 = \min\{\mu_1 + \sigma_1 U_0, U_1\}$ and $X_2 = \min\{\mu_2 +\sigma_2 U_0, U_2\}$, then the joint distribution of $(X_1, X_2)$ is called the \textbf{Marshall-Olkin bivariate Pareto (MOBVPA) distribution} or singular bivariate Pareto distribution. The joint survival function of $(X_1, X_2)$ can be written as;
\begin{align*}
S(x_1, x_2) &= (1 + z)^{-\alpha_0}\Big(1 + \frac{x_1 - \mu_1}{\sigma_1}\Big)^{-\alpha_1}\Big(1 + \frac{x_2 - \mu_2}{\sigma_2}\Big)^{-\alpha_2}\\&=
\begin{cases}
S_1(x_1, x_2), \quad \text{if $\frac{x_1 - \mu_1}{\sigma_1}$ \textless $\frac{x_2 - \mu_2}{\sigma_2}$}
\\
S_2(x_1, x_2), \quad \text{if $\frac{x_1 - \mu_1}{\sigma_1}$ \textgreater $\frac{x_2 - \mu_2}{\sigma_2}$}\\
S_{0}(x), \quad \text{if $\frac{x_1 - \mu_1}{\sigma_1}$ = $\frac{x_2 - \mu_2}{\sigma_2} = x$} 
\end{cases}
\end{align*}
where 
\begin{align*}
S_1(x_1, x_2)&=  \Big(1 + \frac{x_2 - \mu_2}{\sigma_2}\Big)^{-\alpha_0 - \alpha_2} \Big(1 + \frac{x_1 - \mu_1}{\sigma_1}\Big)^{-\alpha_1}\\
S_2(x_1, x_2)&=  \Big(1 + \frac{x_2 - \mu_2}{\sigma_2}\Big)^{-\alpha_2}\Big(1 + \frac{x_1 - \mu_1}{\sigma_1}\Big)^{- \alpha_0 - \alpha_1}\\
S_0(x)&=\Big(1 + x\Big)^{- \alpha_0 - \alpha_1 - \alpha_2}
\end{align*}
so it's pdf that can be written as 
\begin{align*}
f(x_1, x_2) &=
\begin{cases}
f_1(x_1, x_2), \quad \text{if $\frac{x_1 - \mu_1}{\sigma_1}$ \textless $\frac{x_2 - \mu_2}{\sigma_2}$}
\\
f_2(x_1, x_2), \quad \text{if $\frac{x_1 - \mu_1}{\sigma_1}$ \textgreater $\frac{x_2 - \mu_2}{\sigma_2}$}\\
f_{0}(x), \quad \text{if $\frac{x_1 - \mu_1}{\sigma_1}$ = $\frac{x_2 - \mu_2}{\sigma_2} = x$}
\end{cases}
\end{align*}
where 
\begin{align*}
f_1(x_,x_2) &= \frac{\alpha_1 (\alpha_0 + \alpha_2)}{\sigma_1 \sigma_2}\Big(1 + \frac{x_2 - \mu_2}{\sigma_2}\Big)^{-\alpha_0 - \alpha_2 - 1}\Big(1 +\frac{x_1 - \mu_1}{\sigma_1}\Big)^{-\alpha_1 - 1}\\
f_2(x_1, x_2) &=\frac{\alpha_2 (\alpha_0 + \alpha_1)}{\sigma_1 \sigma_2}\Big(1 + \frac{x_2-\mu_2}{\sigma_2}\Big)^{- \alpha_2 - 1}\Big(1 + \frac{x_1 - \mu_1}{\sigma_1}\Big)^{-\alpha_0 - \alpha_1 - 1} \\
f_0(x) &= \alpha_0(1 + x)^{-\alpha_0 - \alpha_1 - \alpha_2 - 1}
\end{align*}

 We denote this distribution as $MOBVPA(\mu_1, \mu_2, \sigma_1, \sigma_2, \alpha_0, \alpha_1, \alpha_2)$.  In this paper we choose $\mu_1 = \mu_2 = 0$ and $\sigma_1 = \sigma_2 = 1$. Then the joint PDF is

\begin{align}
f(x_1, x_2) &= \begin{cases}
\alpha_1(\alpha_0 + \alpha_2)(1 + x_1)^{-\alpha_1 - 1}(1 + x_2)^{-\alpha_0 - \alpha_2 - 1}, \quad \text{if $x_1$ \textless $x_2$}\\
\alpha_2(\alpha_0 + \alpha_1)(1 + x_2)^{-\alpha_2 - 1}(1 + x_1)^{-\alpha_0 - \alpha_1 - 1}, \quad \text{if $x_1$ \textgreater $x_2$}\\
\alpha_0(1 + x)^{-\alpha_0 - \alpha_1 - \alpha_2 - 1}, \quad \text{if $x_1 = x_2 = x$}
\end{cases}  \label{01e1}
\end{align}

\subsection{Likelihood Function}

  The likelihood function corresponding to this pdf is given by,
\begin{eqnarray}
l(x_1, x_2; \alpha_0, \alpha_1, \alpha_2) &=& \alpha_0^{n_0}\alpha_1^{n_1}\alpha_2^{n_2}(\alpha_0 + \alpha_1)^{n_2}\nonumber\\ &&(\alpha_0 + \alpha_2)^{n_1}\prod_{i\in I_0}(1 + x_{1i})^{-(\alpha_0 + \alpha_1 + \alpha_2 - 1)} \prod_{i\in I_1}(1 + x_{1i})^{-\alpha_1 - 1}\nonumber\\&&(1 + x_{2i})^{-\alpha_0 - \alpha_2 - 1}
\prod_{i \in I_2}(1 + x_{1i})^{-\alpha_0 - \alpha_1 - 1}(1 + x_{2i})^{-\alpha_2 - 1}  \label{4e1}
\end{eqnarray}

where $I_0 = \{(x_1, x_2) \mid x_1 = x_2\}$, $I_1 = \{(x_1, x_2) \mid x_1 < x_2\}$ and $I_2 = \{(x_1, x_2) \mid x_1 > x_2\}$. 

Therefore log-likelihood function takes the form,
\begin{eqnarray*}
&& L(\alpha_{0}, \alpha_{1}, \alpha_{2})\\ & = & n_{1}\ln\alpha_{1} + n_{1}\ln(\alpha_{0} + \alpha_{2}) - (\alpha_{1} + \alpha_{2} + 1)\sum_{i \in I_{1}}^{n} \ln(1 + x_{2i})\\ & - & (\alpha_{1} + 1)\sum_{i \in I_{1}}^{} \ln(1 + x_{1i}) + n_{2}\ln\alpha_{2} + n_{2}\ln(\alpha_{0} + \alpha_{1})\\ & - & (\alpha_{0} + \alpha_{1} + 1)\sum_{i \in I_{2}}^{n} \ln (1 + x_{1i}) - (\alpha_{2} + 1)\sum_{i \in I_{2}}^{n} \ln(1 + x_{2i})\\ & + & n_{0}\ln\alpha_{0} - (\alpha_{0} + \alpha_{1} + \alpha_{2} + 1)\sum_{i \in I_{0}}^{n} \ln(1 + x_{1i})
\end{eqnarray*}

\subsection{Prior Assumption}

\subsubsection{Gamma Prior}

 We assume that $\alpha_0$, $\alpha_1$, and $\alpha_2$ are distributed according to the gamma distribution with shape parameters $k_i$ and scale parameters $\theta_i$, i.e.,
\begin{eqnarray}
\alpha_0 \sim \Gamma(k_0, \theta_0) \equiv \mathrm{Gamma}(k_0, \theta_0) \nonumber \\
\alpha_1 \sim \Gamma(k_1, \theta_1) \equiv \mathrm{Gamma}(k_1, \theta_1) \nonumber \\
\alpha_2 \sim \Gamma(k_2, \theta_2) \equiv \mathrm{Gamma}(k_2, \theta_2)
\end{eqnarray}
The probability density function of the Gamma Distribution is given by,
\begin{eqnarray}
f_{\Gamma}(x; k, \theta) = \frac{1}{\Gamma(k)\theta^{k}}x^{k - 1}e^{-\frac{x}{\theta}}
\end{eqnarray} 
Here $\Gamma(k)$ is the gamma function evaluated at $k$.

\subsubsection{Reference Prior Assumption}

  We calculate the expression using Bernardo's reference Prior [\cite{BergerBernardo:1992}, \cite{Bernardo:1979}] in this context.  The following priors are applicable in finding directly the conditional posterior distribution of one parameter given the others and the data.  We use conditional prior of one parameter given the others instead of proposing the unconditional ones. Since we are planning to use Slice cum Gibbs sampler, we do not need the expression of full posterior distribution. Writing the joint unconditional prior will lead to a very complicated expression.  We avoid the same and directly write the conditional distribution of one parameter given the others.        

The expressions are as follows : 
\begin{eqnarray*}
P_0 = \pi(\alpha_{0} | \alpha_{1}, \alpha_{2}) & \propto &  \sqrt{-\left(\frac{\partial^{2} L}{\partial \alpha^{2}_{0}}\right)}\\
           &  = & \sqrt{\frac{n_{0}}{(\alpha_{0})^{2}} + \frac{n_{2}}{(\alpha_{0} + \alpha_{1})^{2}} + \frac{n_{1}}{(\alpha_{0} + \alpha_{2})^{2}}}         
\end{eqnarray*}
\begin{eqnarray*}
P_1 = \pi(\alpha_{1} | \alpha_{0}, \alpha_{2}) & \propto & \sqrt{-\left(\frac{\partial^{2} L}{\partial \alpha^{2}_{1}}\right)}
           = \sqrt{\frac{n_{1}}{\alpha^{2}_{1}} + \frac{n_{2}}{(\alpha_{0} + \alpha_{1})^{2}}}          
\end{eqnarray*}
\begin{eqnarray*}
P_2 = \pi(\alpha_{2} | \alpha_{0}, \alpha_{1}) & \propto & \sqrt{-\left(\frac{\partial^{2} L}{\partial \alpha^{2}_{2}}\right)}
            = \sqrt{\frac{n_{2}}{\alpha^{2}_{2}} + \frac{n_{1}}{(\alpha_{0} + \alpha_{2})^{2}}}          
\end{eqnarray*}

\subsection{Bayes Estimates}

  In this section we provide the bayes estimates of the unknown parameters namely $\alpha_0, \alpha_1$, and $\alpha_2$ for singular bivariate Pareto distribution using Lindley approximation and Slice cum Gibbs Sampler method.  In this paper we use step-out slice sampling as described by \cite{Neal:2003}.  We can provide the expression of full posterior when posterior is constructed based on Gamma prior.  The full posterior of $(\alpha_0, \alpha_1, \alpha_2)$ given the data $D_2$ based on the gamma prior $\pi(\cdot)$ is,
\begin{align}
\pi(\alpha_0, \alpha_1, \alpha_2 | D_2)\propto & \quad l(\alpha_0, \alpha_1, \alpha_2 | D_2)\pi(\alpha_0, \alpha_1, \alpha_2) \nonumber\\ = &
\alpha_0^{n_0}\alpha_1^{n_1}\alpha_2^{n_2}(\alpha_0 + \alpha_1)^{n_2}(\alpha_0 + \alpha_2)^{n_1} \nonumber\\ &\prod_{i\in I_0}(1 + x_{1i})^{-(\alpha_0 + \alpha_1 + \alpha_2-1)} \prod_{i\in I_1}(1 + x_{1i})^{-\alpha_1 - 1} \nonumber\\&(1 + x_{2i})^{-\alpha_0 - \alpha_2 - 1}
\prod_{i \in I_2}(1 + x_{1i})^{-\alpha_0 - \alpha_1 - 1}(1 + x_{2i})^{-\alpha_2 - 1} \nonumber\\&\times 
\alpha_0^{k_0 - 1}\alpha_1^{k_1 - 1}\alpha_2^{k_2 - 1}e^{-(\frac{\alpha_0}{\theta_0} + \frac{\alpha_1}{\theta_1} + \frac{\alpha_2}{\theta_2})} \nonumber\\=&
\pi_1(\alpha_0, \alpha_1, \alpha_2 | D_2)\quad \text{(say)} \label{4e2}
\end{align}

 Therefore, if we want to compute the bayes estimate of some function of $\alpha_0$, $\alpha_1$  and $\alpha_2$, say $ g(\alpha_0, \alpha_1,\alpha_2)$, the bayes estimate of $g$, say $\hat{g}$ under the squared error loss function is the posterior mean of $g$, i.e.
\begin{equation}
\hat{g} = \frac{\int_{0}^{\infty}\int_{0}^{\infty}\int_{0}^{\infty}g(\alpha_0, \alpha_1, \alpha_2)\pi_1(\alpha_0, \alpha_1, \alpha_2 | D_2)d\alpha_0 d\alpha_1 d\alpha_2}{\int_{0}^{\infty}\int_{0}^{\infty}\int_{0}^{\infty}\pi_1(\alpha_0, \alpha_1, \alpha_2|D_2)d\alpha_0 d\alpha_1 d\alpha_2} \label{ghat}
\end{equation}

\subsection{The Full log- conditional posterior distributions in Gamma prior and Reference Prior}

\begin{eqnarray*}
&&\ln(\pi(\alpha_0 \mid \alpha_1, \alpha_2, x_1, x_2))\\ & = & n_0\log(\alpha_0) + n_2\log(\alpha_0 + \alpha_1) + n_1\log(\alpha_0 + \alpha_2)\\ & - & (\alpha_0 + \alpha_1 + \alpha_2 - 1)\sum_{i \in I_{0}}^{} \log(1 + x_{1i}) - (\alpha_0 + \alpha_2 + 1)\sum_{i \in I_{1}}^{} \log(1 + x_{2i})
\\ & - & (\alpha_0 + \alpha_1 + 1)\sum_{i \in I_{2}}^{} \log(1 + x_{1i}) + (k_0 - 1)\ln\alpha_0 - \frac{\alpha_0}{\theta_0}
\end{eqnarray*}
 \begin{eqnarray*}
&&\ln(\pi(\alpha_1 \mid \alpha_0, \alpha_2, x_1, x_2))\\ & = & n_1\log(\alpha_1) + n_2\log(\alpha_0 + \alpha_1) - (\alpha_0 + \alpha_1 + \alpha_2 - 1)\sum_{i \in I_{0}}^{}\log(1 + x_{1i})\\ & - & (\alpha_1 + 1)\sum_{i \in I_{1}}^{}\log(1 + x_{1i}) - (\alpha_0 + \alpha_1 + 1)\sum_{i \in I_{2}}^{} \log(1 + x_{1i})\\ & + & (k_1 - 1)\ln\alpha_1 - \frac{\alpha_1}{\theta_1}
\end{eqnarray*}
\begin{eqnarray*}
&&\ln(\pi(\alpha_2 \mid \alpha_0, \alpha_1, x_1, x_2))\\ & = & n_2\log(\alpha_2) + n_1\log(\alpha_0 + \alpha_2) - (\alpha_0 + \alpha_1 + \alpha_2- 1)\sum_{i \in I_{0}}^{}\log(1 + x_{1i})\\ & - & (\alpha_0 + \alpha_2 + 1)\sum_{i \in I_{1}}^{}\log(1 + x_{2i}) - (\alpha_2 + 1)\sum_{i \in I_{2}}^{} \log(1 + x_{2i})\\ & + & (k_2 - 1)\ln\alpha_2 - \frac{\alpha_2}{\theta_2}
\end{eqnarray*}

  We use conditional prior of one parameter given the others instead of proposing the unconditional ones.  Since we are planning to use Slice cum Gibbs sampler, we do not need the expression of full posterior distribution.  Writing the joint unconditional prior will lead to a very complicated expression.  We avoid the same and directly write the conditional distribution of one parameter given the others.  The expressions are as follows :

\begin{eqnarray*}
&&\ln(\pi(\alpha_0 \mid \alpha_1, \alpha_2, x_1, x_2))\\ & = & n_0\log(\alpha_0) + n_2\log(\alpha_0 + \alpha_1) + n_1\log(\alpha_0 + \alpha_2)\\ & - & (\alpha_0 + \alpha_1 + \alpha_2 - 1)\sum_{i \in I_{0}}^{} \log(1 + x_{1i}) - (\alpha_0 + \alpha_2 + 1)\sum_{i \in I_{1}}^{} \log(1 + x_{2i})
\\ & - & (\alpha_0 + \alpha_1 + 1)\sum_{i \in I_{2}}^{} \log(1 + x_{1i}) + \log(P_0);
\end{eqnarray*}
 \begin{eqnarray*}
&&\ln(\pi(\alpha_1 \mid \alpha_0, \alpha_2, x_1, x_2))\\ & = & n_1\log(\alpha_1) + n_2\log(\alpha_0 + \alpha_1) - (\alpha_0 + \alpha_1 + \alpha_2 - 1)\sum_{i \in I_{0}}^{}\log(1 + x_{1i})\\ & - & (\alpha_1 + 1)\sum_{i \in I_{1}}^{}\log(1 + x_{1i}) - (\alpha_0 + \alpha_1 + 1)\sum_{i \in I_{2}}^{} \log(1 + x_{1i}) + \log(P_1);
\end{eqnarray*}
\begin{eqnarray*}
&&\ln(\pi(\alpha_2 \mid \alpha_0, \alpha_1, x_1, x_2))\\ & = & n_2\log(\alpha_2) + n_1\log(\alpha_0 + \alpha_2) - (\alpha_0 + \alpha_1 + \alpha_2- 1)\sum_{i \in I_{0}}^{}\log(1 + x_{1i})\\ & - & (\alpha_0 + \alpha_2 + 1)\sum_{i \in I_{1}}^{}\log(1 + x_{2i}) - (\alpha_2 + 1)\sum_{i \in I_{2}}^{} \log(1 + x_{2i}) + \log(P_2);
\end{eqnarray*}

\subsection{General Lindley Approximation}

  We use Lindley Approximation (\cite{Lindley:1980}) technique to approximate $(\ref{ghat})$ which is same as approximate evaluation of integral of the form :
\begin{equation}
\frac{\int w(\theta)e^{M(\theta)}d\theta}{\int v(\theta)e^{M(\theta)}d\theta}  \label{3e1}
\end{equation}
where $\theta = (\theta_1, \theta_2, \theta_3, \cdots, \theta_k)$ is a parameter. Here $w(\theta)$, $v(\theta)$ and $M(\theta)$ are any arbitrary functions of $\theta$.  
 
  Let us consider $\textbf{x}$ as a sample of size n taken from a population with probability density function $f(x | \theta)$ and X be the corresponding random variable.  Let's denote the likelihood function as $l(\theta | \textbf{x})$ and log-likelihood function as $ L(\theta | \textbf{x})$. 

  We assume that $\pi(\theta)$ is a prior distribution of $\theta$ and $g(\theta)$ is any arbitrary function of $\theta$. Under squared error loss function, the bayes estimate of $g(\theta)$ is the posterior mean of $g(\theta)$. Then the Bayes estimate of $g(\theta)$ is,
\begin{align}
\hat{g}_B =& \frac{\int_{\theta}g(\theta)l(\theta|\textbf{x})\pi(\theta)d\theta}{\int_{\theta}l(\theta|\textbf{x})\pi(\theta)d\theta} \label{3e3}
\end{align}

  Let us assume that  $\rho(\theta) = \log \pi(\theta)$. So equation $(\ref{3e3})$ can be written as, 
\begin{equation}
\hat{g}_B = \frac{\int_{\theta} g(\theta)e^{[L(\theta|\textbf{x}) + \rho(\theta)]}d\theta}{\int_{\theta}e^{[L(\theta|\textbf{x}) + \rho(\theta)]}d\theta}  \label{3e4}
\end{equation}

  In this case  $v(\theta) = \pi(\theta)$, $w(\theta) = g(\theta)\pi(\theta)$ and $M(\theta) = L(\theta|\textbf{x})$.

 After simplification we can write the equation (\ref{3e4}) as
\begin{equation}
\hat{g}_B = g + \frac{1}{2}\sum(g_{ij} + 2g_{i}\rho_{j})\sigma_{ij} + \frac{1}{2}\sum L_{ijk}g_l\sigma_{ij}\sigma_{kl} \label{3e5}
\end{equation}
where $i, j, k, l = 1, 2, 3, \cdots, k$. Many partial derivatives occur in RHS of the equation ($\ref{3e5}$).  Here $L_{ijk}$ is the third order partial derivative with respect to $\alpha_i, \alpha_j, \alpha_k$, whereas $g_{i}$ is the first order partial derivative with respect to $\alpha_i$ and $g_{ij}$ is the second order derivative with respect to $\alpha_i$ and $\alpha_j$.  We denote $\sigma_{ij}$ as the $(i, j)$-th element of the inverse of the matrix $\{ L_{ij} \}$. All term in right hand side of the equation ($\ref{3e5}$) are calculated at MLE of $\theta$ $(=\hat{\theta}, \text{say})$.

\subsection{Lindley Approximation for 3-Parameter singular MOBVPA:}

  Let $\alpha_0, \alpha_1, \alpha_2$ be the parameters of corresponding distribution and $\pi(\alpha_0, \alpha_1, \alpha_2)$ is the joint prior distribution of $\alpha_0, \alpha_1$ and $\alpha_2$. Then the bayes estimate of any function of $\alpha_0$, $\alpha_1$ and $\alpha_2$, say $g = g(\alpha_0, \alpha_1, \alpha_2)$ under the squared error loss function is,
\begin{align}
\hat{g}_B = &\frac{\int_{(\alpha_0, \alpha_1, \alpha_2)}g(\alpha_0, \alpha_1, \alpha_2)e^{[L(\alpha_0, \alpha_1, \alpha_2) + \rho(\alpha_0,\alpha_1, \alpha_2)]}d(\alpha_0, \alpha_1, \alpha_2)}{\int_{(\alpha_0, \alpha_1, \alpha_2)}e^{[L(\alpha_0, \alpha_1, \alpha_2) + \rho(\alpha_0,\alpha_1, \alpha_2)]}d(\alpha_0, \alpha_1, \alpha_2)}   \label{bayes3e6}
\end{align}

 where $L(\alpha_0, \alpha_1, \alpha_2)$ is  log-likelihood function and $\rho(\alpha_0, \alpha_1, \alpha_2)$ is logarithm of joint prior of $\alpha_0, \alpha_1$ and $\alpha_2$ i.e $\rho(\alpha_0, \alpha_1, \alpha_2) = \log \pi(\alpha_0, \alpha_1, \alpha_2)$. By the Lindley approximation, (\ref{Lindley:1980}) can be written as,

\begin{align*}
\hat{g}_B = &g(\hat{\alpha}_0, \hat{\alpha}_1, \hat{\alpha}_2) + (g_0b_0 + g_1b_1 + g_2b_2 + b_3 + b_4) + \frac{1}{2}[A(g_0\sigma_{00} + g_1\sigma_{01} + g_2\sigma_{02})\\& + B(g_0\sigma_{10} + g_1\sigma_{11} + g_2\sigma_{12}) + C(g_0\sigma_{20} + g_1\sigma{21} + g_2\sigma_{22})]
\end{align*} 
where $\hat{\alpha}_0, \hat{\alpha}_1$ and $\hat{\alpha}_2$ are the MLE of $\alpha_0, \alpha_1$ and $\alpha_2$ respectively.

\begin{align*}
b_i =& \rho_0\sigma_{i0} + \rho_1\sigma_{i1} + \rho_2\sigma_{i2}, \quad
 i = 0, 1, 2\\ 
 b_3 = & g_{01}\sigma_{01} + g_{02}\sigma_{02} + g_{12}\sigma_{12}\\
 b_4 = & \frac{1}{2}(g_{00}\sigma_{00} + g_{11}\sigma_{11} + g_{22}\sigma_{22})\\
 A = &\sigma_{00}L_{000} + 2\sigma_{01}L_{010} + 2\sigma_{02}L_{020} + 2\sigma_{12}L_{120} + \sigma_{11}L_{110} + \sigma_{22}L_{220}\\
 B = &\sigma_{00}L_{001} + 2\sigma_{01}L_{011} + 2\sigma_{02}L_{021} + 2\sigma_{12}L_{121} + \sigma_{11}L_{111} + \sigma_{22}L_{221}\\
 C = &\sigma_{00}L_{002} + 2\sigma_{01}L_{012} + 2\sigma_{02}L_{022} + 2\sigma_{12}L_{122} + \sigma_{11}L_{113} + \sigma_{22}L_{222}
\end{align*}

Also \begin{equation*}
\rho_i = \bigg[\frac{\delta \rho}{\delta \alpha_i}\bigg]_{\text{at} (\hat{\alpha}_0, \hat{\alpha}_1, \hat{\alpha}_2)}, \quad g_i = \bigg[\frac{\delta g(\alpha_0, \alpha_1, \alpha_2)}{\delta\alpha_i}\bigg]_{\text{at} (\hat{\alpha}_0, \hat{\alpha}_1, \hat{\alpha}_2)}, \quad i = 0, 1, 2
\end{equation*}
\begin{equation*}
g_{ij}=\bigg[\frac{\delta^2 g(\alpha_0, \alpha_1, \alpha_2)}{\delta\alpha_i\delta\alpha_j}\bigg]_{\text{at} (\hat{\alpha}_0, \hat{\alpha}_1, \hat{\alpha}_2)}, \quad L_{ij}=\bigg[\frac{\delta^2L(\alpha_0, \alpha_1, \alpha_2)}{\delta\alpha_i\delta\alpha_j}\bigg]_{\text{at} (\hat{\alpha}_0, \hat{\alpha}_1, \hat{\alpha}_2)}, i = 0, 1, 2 
\end{equation*}
\begin{equation*}
L_{ijk}=\bigg[\frac{\delta^3 L(\alpha_0, \alpha_1, \alpha_2)}{\delta\alpha_i\delta\alpha_j\delta\alpha_k}\bigg]_{\text{at} (\hat{\alpha}_0, \hat{\alpha}_1, \hat{\alpha}_2)} i = 0, 1, 2
\end{equation*} 

 Here $\sigma_{ij}$ is the $(i, j)-th$ element  of the inverse of the matrix $\{ L_{ij} \}$ all evaluted at the MLE of $\alpha_0, \alpha_1$ and $\alpha_2$ i.e at $(\hat{\alpha}_0, \hat{\alpha}_1, \hat{\alpha}_2)$.  Now $\rho = \log \pi(\alpha_0, \alpha_1, \alpha_2)$ then,
$ \rho_0 =  \frac{k_0 - 1}{\alpha_0} - \frac{1}{\theta_0} $, $ \rho_1 = \frac{k_1 - 1}{\alpha_1} - \frac{1}{\theta_1} $, $\rho_2 = \frac{k_2 - 1}{\alpha_2} - \frac{1}{\theta_2}.$
\begin{align*}
L_{00} = & -\frac{n_2}{(\hat{\alpha}_0 + \hat{\alpha}_1)^2} - \frac{n_1}{(\hat{\alpha}_0 + \hat{\alpha}_2)^2} - \frac{n_0}{\alpha_0^2}\\
L_{11} = & -\frac{n_1}{(\hat{\alpha}_1)^2} - \frac{n_2}{(\hat{\alpha}_0 + \hat{\alpha}_1)^2}\\
L_{22} = & -\frac{n_1}{(\hat{\alpha}_0 + \hat{\alpha}_2)^2}\\
L_{01} = & -\frac{n_2}{(\hat{\alpha}_0 + \hat{\alpha}_1)^2} = L_{10}\\
L_{02} = & -\frac{n_1}{(\hat{\alpha}_0 + \hat{\alpha}_2)^2} = L_{20}\\
L_{12} = &0 = L_{21}
\end{align*}
the values of $L_{ijk}$ for $i, j, k = 0, 1, 2$ are given by
\begin{align*}
L_{000} = & \frac{2n_2}{(\hat{\alpha}_0 + \hat{\alpha}_1)^3} + \frac{2n_1}{(\hat{\alpha}_0 + \hat{\alpha}_2)^3} + \frac{3n_0}{\alpha_0}\\
L_{111} = & \frac{2n_1}{(\hat{\alpha}_1)^3} + \frac{2n_2}{(\hat{\alpha}_0 + \hat{\alpha}_1)^3}\\
L_{222} = & \frac{2n_2}{(\hat{\alpha}_2)^3} + \frac{2n_1}{(\hat{\alpha}_0 + \hat{\alpha}_2)^3}\\
L_{001} = & \frac{2n_2}{(\hat{\alpha}_0 + \hat{\alpha}_1)^3} = L_{010} = L_{100}\\
L_{002} = & \frac{2n_1}{(\hat{\alpha}_0 + \hat{\alpha}_2)^3} = L_{020} = L_{200}\\
L_{011} = & \frac{2n_2}{(\hat{\alpha}_0 + \hat{\alpha}_1)^3} = L_{101} = L_{110}\\
L_{012} = & 0 = L_{021} = L_{102} = L_{120} = L_{201} = L_{210}\\
L_{022} = & \frac{2n_1}{(\hat{\alpha}_0 + \hat{\alpha}_2)^3} = L_{202} = L_{220}\\
L_{112} = & 0 = L_{121} = L_{211}\\
L_{122} = & 0 = L_{212} = L_{221}
\end{align*}

 Now we can obtain the Bayes estimates of $\alpha_0, \alpha_1$ and $\alpha_2$ under squared error loss function.

\begin{enumerate}

\item[(i)] For $\alpha_0$, choose $g(\alpha_0, \alpha_1, \alpha_2) = \alpha_0$. 
So Bayes estimates of $\alpha_0$ can be written as,
\begin{equation}
\hat{\alpha}_{0B} = \hat{\alpha}_0 + b_0 + \frac{1}{2}[A\sigma_{00} + B\sigma_{10} + C\sigma_{20}]
\end{equation}

\item[(ii)] For $\alpha_1$, choose $g(\alpha_0, \alpha_1, \alpha_2) = \alpha_1$. So bayes estimates of $\alpha_1$ can be written as,
\begin{equation}
\hat{\alpha}_{1B} = \hat{\alpha}_1 + b_1 + \frac{1}{2}[A\sigma_{01} + B\sigma_{11} + C\sigma_{21}]
\end{equation}

\item[(iii)] For $\alpha_2$, choose $g(\alpha_0, \alpha_1, \alpha_2) = \alpha_2$. So bayes estimates of $\alpha_2$ can be written as,
\begin{equation}
\hat{\alpha}_{2B} = \hat{\alpha}_2 + b_2 + \frac{1}{2}[A\sigma_{02} + B\sigma_{12} + C\sigma_{22}]
\end{equation}

\end{enumerate}

\noindent{Remark :}  We replace MLE by its estimates obtained through EM algorithm \cite{DeyPaul:2017} while calculating the Lindley approximation. 
 
\section{Constructing credible Intervals for $\underline{\theta}$}

 We find the credible intervals for parameters as described by Chen and Shao \cite{ChenShao:1999}.  Let assume $\underline{\theta}$ is vector.  To obtain credible intervals of first variable $\theta_{1i}$, we order $\{\theta_{1i}\}$, as $\theta_{1(1)} < \theta_{1(2)} < \cdots  < \theta_{1(M)}$. Then 100(1 - $\gamma$)$\%$ credible interval of $\theta_1$ become
$$(\theta_{1(j)}, \theta_{1(j + M - M\gamma)}), \qquad for \quad j = 1, \cdots , M\gamma$$

Therefore 100(1 - $\gamma$)$\%$ credible interval for $\theta_1$ becomes $(\theta_{1(j^*)}, \theta_{1(j^* + M - M\gamma)}),$ where $j^*$ is such that $$\theta_{1(j^* + M - M\gamma)} - \theta_{1(j^*)} \leq \theta_{1(j + M - M\gamma)} - \theta_{1(j)}$$
for all $j = 1, \cdots , M\gamma$. Similarly, we can obtain the credible interval for other co-ordinates of $\theta$.

  We have scope to construct such intervals when full posterior is not known and tractable.  In this paper we calculate the bayesian confidence interval for both gamma prior and reference prior.  We skip working with full expression of posterior under reference prior as it is not tractable.  We use R package coda to obtain the credible intervals described above. 

\section{Numerical Results}

  The numerical results are obtained by using package R 3.2.3. The codes are run at IIT Guwahati computers with model : Intel(R) Core(TM) i5-6200U CPU \@ 2.30GHz. The codes will be available on request to authors.   

 We use the following hyper parameters of prior as gamma : $k_0 = 2,\ \theta_0 = 3,\ k_1 = 4,\ \theta_1 = 3,\ k_2 = 3,\ \theta_2 = 2$.  Bayes estimates, mean square errors, credible intervals are calculated for all the parameters $\alpha_0$, $\alpha_1$ and $\alpha_2$ using both gamma prior and reference prior.  Table-\ref{tab-sinbvpa1} and Table-\ref{tab-sinbvpa2} show the results obtained by different methods, e.g. Lindley and Slice sampling etc with different set of priors like Gamma and reference for two different parameter sets.  In slice cum gibbs sampling we take burn in period as 500.  Bayes estimates are calculated based on 2000 and more iterations after burn-in period.  We make further investigation on sample size needed for all the methods to work. We observe that Slice-cum gamma works even for a sample size like 50 for small parameter values in case of singular MOBVPA.  When original sample is drawn from parameters little bigger, sample size needed to converge the algorithm becomes more.  Slice-cum Gibbs with reference prior as prior requires slightly more sample size like 250 or more to converge.  However Lindley approximation works for sample size around 150 in almost all cases. 

\section{Data Analysis}  We study the two data sets used in two previous papers \cite{DeyPaul:2017}.  This data set is used to model singular Marshall-Olkin bivariate Pareto distribution.  We get the estimates of parameters through EM algorithm for singular Marshall-Olkin bivariate Pareto distribution as $\mu_1 = 0.0158$, $\mu_2 = 0.0012$, $\sigma_1 = 3.0647$, $\sigma_2 = 1.9631$, $\alpha_0 = 2.5251$, $\alpha_1 = 1.028$, $\alpha_2 = 1.4758$.  The paper deals with three parameter set up.  Since direct real life data which will model three parameter MOBVPA is not available.  Therefore we modify the data with location and scale transformation.  This transformation will affect cardinalities of $I_{0}$, $I_{1}$ and $I_{2}$ and thereby the value of likelihood function in singular MOBVPA significantly.  Therefore we modified the algorithm by making a suitable approximation of the number of observations in each of $I_{0}$, $I_{1}$ and $I_{2}$ while calculating the value of likelihood function. We replace $n_{0}$, $n_1$ and $n_2$, the cardinality of cells $I_{0}$, $I_{1}$ and $I_{2}$ by $\tilde{n_{0}}$, $\tilde{n_{1}}$ and $\tilde{n_{2}}$ where $\tilde{n_{i}} = (n_{0} + n_1 + n_2)\frac{\alpha_i}{(\alpha_0 + \alpha_1 + \alpha_2)}$ for $i = 0, 1, 2$. This approximation can be obtained by using the distribution of unknown random cardinalities as multinomial distribution with parameter $(n_{0} + n_{1} + n_{2})$  and $\frac{\alpha_i}{(\alpha_0 + \alpha_1 + \alpha_2)}$ for $i = 0, 1, 2$.  Bayes estimates and credible intervals are calculated and provided in Table-\ref{tab-datasinbvpa}.

\begin{table}[ht!]
{\begin{tabular}[l]{@{}lccccc}\hline
Slice-cum-Gibbs & & & \\ \hline
Gamma Prior & & & \\ \hline
 Parameter Sets & $\alpha_0$ & $\alpha_1$ & $\alpha_2$ \\ 
 Bayes Estimates & 0.7267 & 0.8661 & 1.0207 \\ 
 Credible Intervals & [0.5970, 0.8861] & [0.6936, 1.0239] & [0.8279, 1.1955] \\ \hline
Reference Prior & & & \\ \hline
 Parameter Sets & $\alpha_0$ & $\alpha_1$ & $\alpha_2$ \\ 
 Bayes Estimates & 0.7567 & 0.8128 & 0.9772 \\ 
 Credible Intervals & [0.6155, 0.9206] & [0.6366, 0.9763] & [0.7946, 1.1666] \\ \hline
Lindley &  & & \\ \hline
 Gamma Prior & & & \\ \hline
Original Parameter Sets & $\alpha_0$ & $\alpha_1$ & $\alpha_2$ \\ 
 Bayes Estimates & 0.7547 & 0.8152 & 0.9812 \\ 
\end{tabular}}   
\caption{The Bayes Estimates (BE) and credible interval of singular Marshall-Olkin bivariate Pareto distribution}
\label{tab-dataabsbvpa} 
\end{table}
        
\section{Conclusion} Bayes estimates of the parameters of singular bivariate Pareto under square error loss are obtained both using Lindley and Slice cum Gibbs sampler approach.  Both the methods are working quite well even for moderately large sample size. In case of singular MOBVPA the algorithms work even for small sample size like 50.  Use of informative prior like Gamma and non-informative prior like reference prior is studied in this context.  Posterior using full reference prior requires more attention.  The same study can be made using many other algorithms like importance sampling, HMC etc. This study can be used to find out bayes factor between two or more bivariate distributions which can be an appropriate criteria for discriminating two or more higher dimensional distributions. More work is needed in this direction. The work is in progress.

\bibliographystyle{apa}

\bibliography{bvpa_bayes}

\begin{table}[ht!]
\begin{small}
{\begin{tabular}[l]{@{}lccccc}\hline
Slice-cum-Gibbs & & & \\ \hline
 Gamma Prior & & & \\ \hline
 n = 450 & & & \\ \hline     
 Original Parameter Sets & $\alpha_0 = 0.1$ & $\alpha_1 = 0.2$ & $\alpha_2 = 0.4$ \\
  Starting Value & 0.4165 & 0.7933 & 0.8250 \\ 
  Bayes Estimates & 0.0768 & 0.2056 & 0.4314 \\ 
  Mean Square Error & 0.0008 & 0.0005 & 0.0017 \\              
  Credible Intervals & [0.0399, 0.1098] & [0.1652, 0.2463] & [0.3826, 0.4862] \\ \hline
 n = 1000 & & & \\ \hline      
 Original Parameter Sets & $\alpha_0 = 0.1$ & $\alpha_1 = 0.2$ & $\alpha_2 = 0.4$ \\ 
  Starting Value & 0.9295 & 0.9741 & 0.0754 \\ 
  Bayes Estimates & 0.0903 & 0.1957 & 0.4083 \\ 
  Mean Square Error & 0.0003 & 0.0003 & 0.0005 \\              
  Credible Intervals & [0.0602, 0.1218] & [0.1640, 0.2313] & [0.3686, 0.4466] \\ \hline
 Reference Prior  &  &  & \\ \hline
 n = 450 & & & \\ \hline     
 Original Parameter Sets & $\alpha_0 = 0.1$ & $\alpha_1 = 0.2$ & $\alpha_2 = 0.4$ \\ 
 Starting Value & 0.4165 & 0.7933 & 0.8280 \\ 
 Bayes Estimates & 0.0139 & 0.2652 & 0.4920 \\ 
 Mean Square Error & 0.0082 & 0.0052 & 0.0098 \\              
 Credible Intervals & [0.0766, 0.1133] & [0.1888, 0.3065] & [0.4071, 0.5546] \\ \hline
 n = 1000 & & & \\ \hline      
 Original Parameter Sets & $\alpha_0 = 0.1$ & $\alpha_1 = 0.2$ & $\alpha_2 = 0.4$ \\ 
 Starting Value & 0.9295 & 0.9741 & 0.7543 \\ 
 Bayes Estimates & 0.0858 & 0.1982 & 0.4117 \\ 
 Mean Square Error & 0.0004 & 0.0003 & 0.0006 \\              
 Credible Intervals & [0.0559, 0.1153] & [0.1671, 0.2348] & [0.3716, 0.4527] \\ \hline
 Lindley &  & & \\ \hline
 n = 450 & & & \\ \hline
 Gamma Prior & & & \\ \hline
 Original Parameter Sets & $\alpha_0 = 0.1$ & $\alpha_1 = 0.2$ & $\alpha_2 = 0.4$ \\ 
 Bayes Estimates & 0.0979 & 0.2020 & 0.4028 \\ 
 Mean Square Error & 0.0003 & 0.0004 & 0.0008 \\ \hline             
 n = 1000 & & & \\ \hline
 Original Parameter Sets & $\alpha_0 = 0.1$ & $\alpha_1 = 0.2$ & $\alpha_2 = 0.4$ \\ 
 Bayes Estimates & 0.0986 & 0.2015 & 0.4018 \\ 
 Mean Square Error & 0.0002 & 0.0002 & 0.0004 \\              
\end{tabular}}   
\caption{The Bayes Estimates (BE), Mean Square Error (MSE) and credible intervals of singular Marshall-Olkin bivariate Pareto distribution with parameters $\alpha_0 = 0.1$, $\alpha_1 = 0.2$ and $\alpha_2 = 0.4$}
\end{small}
\label{tab-sinbvpa1} 
\end{table}

\begin{table}[ht!]
\begin{small}
{\begin{tabular}[l]{@{}lccccc}\hline
 Slice-cum-Gibbs & & & \\ \hline
 Gamma Prior & & & \\ \hline
 n = 450 & & & \\ \hline
 Original Parameter Sets & $\alpha_0 = 4$ & $\alpha_1 = 5$ & $\alpha_2 = 10$ \\ 
  Starting Value & 0.4165 & 0.7933 & 0.8280 \\
  Bayes Estimates & 3.3395 & 5.1126 & 10.7605 \\ 
  Mean Square Error & 0.7270 & 0.3628 & 1.1435 \\  
  Credible Intervals & [2.2738, 4.3973] & [3.9229, 6.2336] & [9.2716, 12.1742] \\ \hline  
 n = 1000 & & & \\ \hline          
 Original Parameter Sets & $\alpha_0 = 4$ & $\alpha_1 = 5$ & $\alpha_2 = 10$ \\ 
 Starting Value & 0.9295 & 0.9741 & 0.7543 \\
  Bayes Estimates & 3.6521 & 4.9348 & 10.1183 \\ 
  Mean Square Error & 0.2692 & 0.1767 & 0.2784 \\              
  Credible Intervals & [2.8865, 4.4232] & [4.1144, 5.7708] & [9.0258, 11.1130] \\ \hline
 n = 450 & & & \\ \hline
 Reference Prior  &  &  & \\ \hline
 Original Parameter Sets & $\alpha_0 = 4$ & $\alpha_1 = 5$ & $\alpha_2 = 10$ \\ 
 Starting Value & 0.4164 & 0.7933 & 0.8281 \\
 Bayes Estimates & 3.3087 & 5.0801 & 10.9049 \\ 
 Mean Square Error & 0.8817 & 0.4815 & 1.5198 \\              
 Credible Intervals & [2.0785, 4.4884] & [3.6979, 6.3754] & [9.2284, 12.6042] \\ \hline
 n = 1000 & & & \\ \hline
 Original Parameter Sets & $\alpha_0 = 4$ & $\alpha_1 = 5$ & $\alpha_2 = 10$ \\ 
 Starting Value & 0.9295 & 0.9741 & 0.0754 \\
 Bayes Estimates & 3.6926 & 4.8668 & 10.1035 \\ 
 Mean Square Error & 0.2595 & 0.2215 & 0.2930 \\              
 Credible Intervals & [2.5156, 4.4454] & [4.0115, 5.8030] & [9.0183, 11.0898] \\ \hline
 Lindley &  & & \\ \hline
 n = 450 & & & \\ \hline
 Gamma Prior & & & \\ \hline
 Original Parameter Sets & $\alpha_0 = 4$ & $\alpha_1 = 5$ & $\alpha_2 = 10$ \\ 
 Bayes Estimates & 4.0285 & 4.9931 & 10.0566 \\ 
 Mean Square Error & 0.3148 & 0.3710 & 0.6805 \\ \hline
 n = 1000 & & & \\ \hline
 Original Parameter Sets & $\alpha_0 = 4$ & $\alpha_1 = 5$ & $\alpha_2 = 10$ \\ 
 Bayes Estimates & 4.0284 & 4.9931 & 10.0566 \\ 
 Mean Square Error & 0.1350 & 0.1681 & 0.2817 \\
\end{tabular}}   
\caption{The Bayes Estimates (BE), Mean Square Error (MSE) and credible intervals of singular Marshal-Olkin bivariate Pareto distribution with parameters $\alpha_0 = 4$, $\alpha_1 = 5$ and $\alpha_2 = 10$}
\end{small}
\label{tab-sinbvpa2} 
\end{table}

\end{document}